\documentclass[aps,prd,epsf,showpacs,amsmath,amssymb,graphics,
twocolumn,10pt]{revtex4}
\usepackage{graphicx}
\usepackage{dcolumn}
\usepackage{bm}

\newcommand{\be}{\begin{equation}}
\newcommand{\ee}{\end{equation}}
\newcommand{\ba}{\begin{eqnarray}}
\newcommand{\ea}{\end{eqnarray}}
\newcommand{\ban}{\begin{eqnarray*}}
\newcommand{\ean}{\end{eqnarray*}}

\newcommand{\eq}[1]{(\ref{#1})}

\begin{document}

\title{High Energy Particle Collisions in Superspinning Kerr Geometry}

\author{Mandar Patil \footnote{ Electronic address: mandarp@tifr.res.in}
and  Pankaj S. Joshi \footnote{ Electronic address: psj@tifr.res.in}}

\affiliation{Tata Institute of Fundamental Research\\
Homi Bhabha Road, Mumbai 400005, India}


\begin{abstract}
We investigate here the particle acceleration and
high energy collision in the Kerr geometry containing a
naked singularity. We show that the center of mass energy of
collision between two particles, dropped in from a finite but
arbitrarily large distance along the axis of symmetry is
arbitrarily large, provided the deviation of the
angular momentum parameter from the mass is very small for
the Kerr naked singularity. The collisions considered here are
between particles, one of them ingoing and the other one being
initially ingoing but which later emerges as an outgoing
particle, after it suffers a reflection from a spatial
region which has a repulsive gravity in the vicinity of the naked
singularity. High energy collisions take place around a region
which marks a transition between the attractive and repulsive
regimes of gravity. We make a critical comparison between our
results and the BSW acceleration mechanism [M. Banados, J. Silk, and
S. M. West, Phys. Rev. Lett. 103, 111102 (2009).] for
extremal Kerr blackholes, and argue that the scenario we give
here has certain distinct advantages. If compact objects
exist in nature with exterior Kerr superspinning geometry then
such high energy collisions would have a significant impact
on the physical processes occurring in its surrounding and
could possibly lead to their own observational signatures.
As an aside we also suggest a curious Gedanken collider
physics experiment which could in principle be constructed
in this geometry.




\end{abstract}
\maketitle

\section{Introduction}

Various terrestrial particle collider experiments
such as the Large Hadron Collider probe physics upto
10 TeV. This energy scale is almost 15 orders of magnitude
smaller than the Planck scale. Particle physics
models in such very high energy regime remain completely
unexplored and untested by means of any terrestrial
collider physics experiment at the current epoch due to
various limitations of technology available to us.
High precision cosmic microwave background experiments
might shed some light on the new physics at high
energies in near future.

An alternative intriguing possibility to study such
a new physics would be to make use of various naturally
occurring extreme gravity astrophysical objects in our
surrounding universe.
In this spirit, it was suggested recently
\cite{BSW}, that
the blackholes that are either extremal or very
close to being extremal, could be used as particle
accelerators to probe new physics all the way upto Planck
scale. In that case, the particles thrown in from infinity
could interact with divergent center of mass energies
near the event horizon of the extremal blackhole, provided
that certain fine-tuning conditions were imposed on
the angular momentum of one of the colliding particles.

In this work we shall show that the Kerr naked singularities
can as well act as particle accelerators to arbitrarily high
energies in the limit where the deviation of angular momentum
parameter $\bar{a}=\frac{a}{M^2}$ is sufficiently small
from unity. The mechanism we propose
here has a distinct advantage over the blackhole case, necessarily
arising from the absence of an event horizon, and due to
the presence of a repulsive gravity regime in the vicinity of the
naked singularity.

Thus the Kerr naked singularity, if it occurs in nature
provides a suitable environment where ultrarelativistic
collisions could take place at extremely large center of mass energies.
The interaction between the particles in this case could be
dominated by new channels of reactions dictated by the beyond standard
model physics. This could provide an excellent laboratory to study
high energy physics unexplored at various terrestrial accelerator experiments.
Unlike the blackhole case, due to the absence of an event horizon
in these models,
a large fraction of the high energy particles produced in these collisions
would either escape away to infinity and get detected on earth,
or they could interact with the surrounding gas and leave behind
their characteristic astrophysical imprint. This might lead to
either direct or indirect observational signatures of the Kerr
superspinning objects.

We also propose here a rather efficient and economical curious
Gedanken collider experiment in the environment of the naked
singularity along the axis of symmetry of the Kerr geometry.
However, this requires an extreme finetuning of a Kerr spin parameter
close to unity.

Our consideration of Kerr naked singular geometries
is motivated by recent theoretical developments in
string theory which suggest that the timelike naked singularities
are naturally resolved and pathological features associated
with them like causality violation are avoided by high
energy modifications to classical general relativity
\cite{Gimon},
thus transcending the classical cosmic censorship conjecture
\cite{Penrose}.
This is consistent with the thinking that in general the
spacetime singularities will be resolved when we use a correct
quantum gravity theory to study the ultra-high and extreme
gravity phenomena. Earlier, the phenomena of naked
singularity resolution is also reported within the framework
of the loop quantum gravity theory
\cite{Goswami}.

The point here is that, even when the final spacetime
singularity may be resolved by the quantum gravity effects
which will be operative at the Planck length in the vicinity
of the singularity, even then the presence or otherwise of an
event horizon, faraway from the singularity in the spacetime,
will make a very significant difference as far
as the physical effects are concerned within either
the blackhole or naked singularity spacetime geometry.

In general, either a blackhole or a naked singularity
will arise as the endstate of gravitational collapse
of a massive star, which shrinks and collapses continually
under the force of its own gravity towards the end of its life
cycle on exhausting the nuclear fuel within.
The formation of blackholes as well as naked singularity
as an endstate of such a gravitational collapse has been rigorously
investigated in recent years.
There are several spherically symmetric models of matter
collapse with realistic equation of states, which lead to the
formation of naked singularities in a gravitational collapse,
starting from a regular initial data specified on an initial
hypersurface. The genericity as well as stability of
these models have also been investigated
\cite{Joshi}.
Basically, the blackhole or naked singularity formation
as the final state of the star collapse is decided by the matter
initial data in terms of the density and pressure and velocity
profiles of the collapsing matter, and the allowed dynamical
evolutions, as permitted by the Einstein equations.
For example, considering a pressure free dust collapse
evolution, the cloud will evolve to a blackhole with the
spacetime singularity hidden within the event horizon, in
the case when the density is taken to be homogeneous always and
when the velocity profile for the collapsing shells
have a specific fine-tuned form. This is the well-known
model case studied by Oppenheimer and Snyder
\cite{OS}.
On the other hand, when the initial density profile is
allowed to be inhomogeneous, say higher at the center of
the star and decreasing slowly with increasing radius,
a naked singularity develops as the collapse end product
\cite{JD}.
Collapse models with non-vanishing pressures have also
been studied in much detail in past years and it is seen
that it is the initial data and
allowed dynamical evolutions in the above sense that
decide the blackhole or naked singularity final states
for the collapsing matter cloud.

In general, it is expected that a collapsing star would
give rise to either a Kerr blackhole or spinning naked singular
configuration as the endstate of a continual gravitational
collapse.
We note that the formation of either of the above
in a gravitational collapse with angular momentum has not
been demonstrated conclusively.
There has been some progress in this
direction using numerical techniques
\cite{Rezolla}.
There has also been an investigation demonstrating formation of
rotating naked singularities in shell collapse in $2+1$
dimensions
\cite{Mann}.
The stability and genericity of this process is yet to be
explored. The formation of superspinning compact objects,
whose size is much smaller than the gravitational radius,
via accretion onto slowly rotating compact objects,
which possess not only mass and angular momentum but
also an anomalous quadruple moment has also been
demonstrated
\cite{Bambi0}.

On the observational front, the existence of event
horizon of supermassive and stellar mass blackhole
candidates has not been shown conclusively, although it
is widely
believed and most likely that these objects are indeed
Kerr blackholes. The measured values of spin parameter by
X-ray continuum emission line shape fitting method is dependent
on the assumptions
regarding accretion onto these objects, modifying
which could yield inferred spin values larger than
unity
\cite{Narayan}.
Keeping this in mind there has been a significant
effort to design strategies to
distinguish blackholes from naked singularities.
The silhouette of Kerr naked singularity is
shown to differ significantly from the blackhole case
in the gravitational lensing studies
\cite{Bambi1}.
Also, the accretion around naked singularities
is shown to differ significantly from the blackholes
\cite{Bambi2}.

We note that the Kerr naked singularity is not a
unique Vacuum, axisymmetric, asymptotically flat
solution to Einstein equations in the absence of an event
horizon. The most general solution in this case happens to
be the Tomimatsu-Sato spacetime
\cite{Sato}.
In this paper we focus only on the Kerr naked singular
solution for simplicity and clarity. Also the stability of
a Kerr superspinning configuration after the singularity is
resolved by quantum gravitational corrections, would
be an interesting issue for investigation in future.

The next Section II discusses the Kerr geometry in
a proper coordinate setting, and in Section III the geodesic
motion is examined with a particular focus on the axis
of symmetry. Then Sections IV and V discuss the particle
acceleration and a collider thought experiment using
the Kerr naked singularity. We then make a comparison of
the naked singularity case with the blackhole case in
Section VI, and certain possible observational implications
of the Kerr naked singularity as related to the particle
acceleration processes are indicated in Section VII.
Finally Section VIII gives concluding remarks.

\section{Kerr spacetime in Kerr-Schild coordinates}

The Kerr metric
\cite{Kerr},\cite{Carter1},\cite{Carter2} is
characterized by two parameters, namely mass $M$ and
angular momentum per unit mass $a=\frac{J}{M}$. When $a \le M$
the Kerr metric represents a blackhole, whereas $a>M$
stands for a naked singularity without an event horizon.
We focus here on the particles following geodesic
motion along the axis of symmetry of Kerr spacetime with $a>M$.
Thus we use the Kerr-Schild (KS) coordinate system
$\left(t,x,y,z\right)$, which is well-behaved around
axis of symmetry \cite{Carter2},\cite{Schild}.

The Kerr spacetime geometry is given by,
\begin{widetext}
\begin{eqnarray}
ds^2= -dt^2+dx^2+dy^2+dz^2+\frac{2Mr^3}{r^4+a^2z^2}
 \left(dt+\frac{zdz}{r}+ \frac{r\left(xdx+ydy\right)-a
\left(xdy-ydx\right)}{r^2+a^2}\right)^2  \label{KKS}
\end{eqnarray}
\end{widetext}
where $r\left(x,y,z\right)$ is a solution to the equation
\begin{equation}
\nonumber
r^4-\left(x^2+y^2+z^2-a^2\right)r^2-a^2z^2=0
\end{equation}
We make a further coordinate transformation and
introduce a new time coordinate $T(t,x,y,z)$ as,
\begin{equation}
dT=dt-\beta dz \label{T2}
\end{equation}
where
\begin{equation}
\nonumber
\beta=-\frac{\frac{z}{r}\frac{2Mr^3}{r^4+a^2z^2}}
{\left(-1+\frac{2Mr^3}{r^4+a^2z^2}\right)}
\end{equation}

In the new coordinate system $\left(T,x,y,z\right)$,
the Kerr metric can now be written as
\begin{widetext}
\begin{eqnarray}
\nonumber
ds^2= \left(-1+\frac{2Mr^3}{r^4+a^2 z^2}\right)dT^2+
\frac{dz^2}{\left(-1+\frac{2Mr^3}{r^4+a^2 z^2}\right)}
\left(-1+\frac{2Mr^3}{r^4+a^2 z^2}-\frac{z^2}{r^2}
\frac{2Mr^3}{r^4+a^2 r^2}\right)+dx^2+dy^2 \\
+\frac{2Mr^3}{r^4+a^2z^2}
 \left(\frac{r\left(xdx+ydy\right)
-a\left(xdy-ydx\right)}{r^2+a^2} \right)^2
+\frac{4Mr^3 \left(dT+\left(\frac{z}{r}+\beta\right)dz
\right)}{r^4+a^2z^2} \left(\frac{r\left(xdx+ydy\right)-a
\left(xdy-ydx\right)}{r^2+a^2} \right) \label{KMP}
\end{eqnarray}
\end{widetext}
We made this transformation so that in the new
coordinate system, in the $T-z$ plane, the metric has
a vanishing non-diagonal term, and it takes a canonical
form resembling the Schwarzschild metric.
The axis of symmetry in new coordinate system is given by
\begin{equation}
x=y=0, \mid z \mid=r
\label{center}
\end{equation}
The metric to leading order, in the spacetime
region close to the symmetry axis can be written as
\begin{widetext}
\begin{eqnarray}
\label{MC}
\nonumber
ds^2= -\left(1-\frac{2Mz}{z^2+a^2}\right)dT^2+dx^2+dy^2
+\left(1-\frac{2Mz}{z^2+a^2}\right)^{-1}dz^2
\end{eqnarray}
\end{widetext}
which is well behaved and regular metric around
the axis, which we use below.

\section{Geodesic motion along axis of symmetry}

In this section we investigate motion of the particles
following timelike geodesics along the axis
of symmetry, {\i.e.} along $z$-axis.

Since the metric coefficients are independent
of $T$, the spacetime admits a Killing vector field
$\xi=\partial_{T}$. For a particle following geodesic
motion, the quantity,$E=-\xi^{\mu}U_{\mu}$ is then conserved,
$U$ being the velocity of the particle, and $E$
is interpreted as the conserved energy per
unit mass of the particle.
The equation depicting conserved
energy $E=-\xi^{\mu}U_{\mu}$ and the normalization $U^{\mu}U_{\mu}=-1$,
together with \eq{MC} and $U^{\mu}=\left(U^T,0,0,U^z\right)$,
allows components of velocity of the particle to
be written as follows,
\begin{eqnarray}
\label{Ut}
U^T=\frac{E}{f} \\
\label{Uz}
\left(U^{z}\right)^2+f=E^2 \\
\nonumber
U^z=\pm \sqrt{E^2-f}
\end{eqnarray}
where
\begin{equation}
f=\left(1-\frac{2Mz}{z^2+a^2}\right)
\label{f}
\end{equation}
Here $\pm$ correspond to the outgoing and ingoing
geodesics respectively. By analogy in Newtonian mechanics,
the function $f$ in \eq{Uz} can be thought of as an
effective potential for a motion along z-axis.

The effective potential $f$ takes a maximum
value at $z=0$ and as $z\rightarrow \infty$. It takes
a minimum value at an intermediate point $z=a$. The behavior of
effective potential is shown in Fig1.
Maximum and minimum values are given by
\begin{eqnarray}
\nonumber
f_{max}=f(z=0)=f(z \rightarrow \infty)=1 \\
f_{min}=f(z=a)=\left(1-\frac{M}{a}\right)=\epsilon >0
\label{Maxmin}
\end{eqnarray}
Since we are dealing with the Kerr solution which
is a naked singular spacetime, the minimum value of $f$
is strictly larger than zero. The parameter $\epsilon >0$
we have introduced in the above indicates
the deviation of the Kerr from the extremal case
where $a=M, \epsilon=0$.
It follows from \eq{Uz} that the particle with
conserved energy per unit mass $E<1$ will be confined
between the values of $z$ as given by,
\begin{eqnarray}
\nonumber
z_{-}=\frac{M-\sqrt{M^2-\left(1-E^2\right)a^2}}
{1-E^2}\\
z_{+}=\frac{M+\sqrt{M^2-\left(1-E^2\right)a^2}}{1-E^2},
\end{eqnarray}
which are the turning points where $U^{z}=0$.

\begin{figure}
\begin{center}
\includegraphics[width=0.5\textwidth]{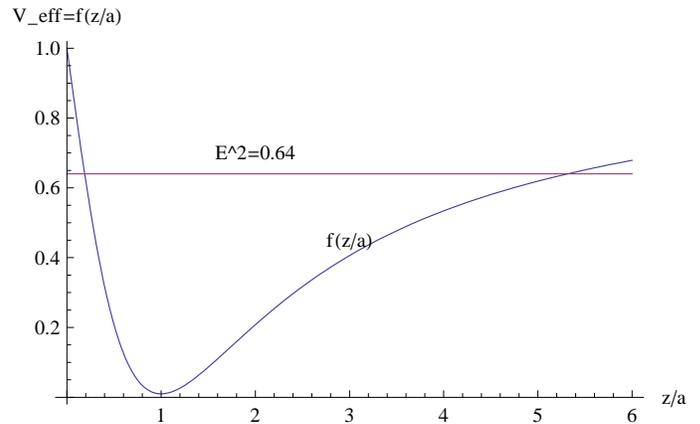}
\caption{\label{fg1}
The effective potential $f$ for a particle traveling along
the axis of symmetry of Kerr spacetime is plotted here against
$\frac{z}{a}$, for a specific angular momentum value
$\frac{a}{M}=1.01$. It is clear from the shape of the effective
potential that the gravity is attractive for $\frac{z}{a}>1.01$,
whereas it is repulsive for $\frac{z}{a}<1.01$.
A test particle with conserved energy per unit mass $E=0.8$,
released from rest at $\frac{z}{a}=5.31$,
initially speeds up upto $\frac{z}{a}=1.01$. It then
slows down and turns back at $\frac{z}{a}=0.19$ as an outgoing
particle. The effective potential admits the minimum at
$\frac{z}{a}=1.01$. Thus a particle with $E=0.3$
will stay at rest at this location.
}
\end{center}
\end{figure}

For an infalling particle, $U^{z}$ goes on
increasing when $z>a$, indicating the attractive nature
of gravity.  The same quantity goes on decreasing
when $z<a$, and eventually it stops and turns back,
thus indicating the 'repulsive nature' of gravity
in this regime. All stationary spacetimes admitting
naked singularities are found to exhibit such a repulsive
gravity effect
in the close neighborhood of singularity
\cite{Felice},\cite{Luongo}.
In Kerr spacetimes, the attractive or repulsive
nature of gravity is roughly determined by whether or
not $\left(r^2-a^2 Sin^2\theta\right)>$ or $ < 0$ respectively
(when expressed in the Boyer-Lindquist coordinates
\cite{Boyer}).
A particle with $E=\sqrt{1-\frac{M^2}{a^2}}$ stays
at rest at $z=a$, which marks a transition between attractive
and repulsive regimes of gravity.

\section{Particle acceleration by Kerr naked singularity}

We consider a collision of two particles,
each of mass $m$ and conserved energy of per unit mass
$E$. One of the particles is taken to be ingoing and
the other one is outgoing.
The center of mass energy $E_{c.m.}$ of
collision between two such particles with velocities
$U^{1},U^{2}$ is given by \cite{BSW},
\begin{equation}
E_{c.m.}^2=2 m^2 \left(1-g_{\mu\nu}U^{1 \mu}U^{2\nu}\right)
\label{Ecm}
\end{equation}
Thus from \eq{Ut},\eq{Uz},\eq{f},\eq{Ecm},
the center of mass energy of collision in this
case would be,
\begin{equation}
E_{c.m.}^2=\frac{4m^2E^2}{f}
\label{Ecm2}
\end{equation}
>From \eq{Maxmin} it can be seen that the
center of mass energy will be maximum if the collision
happens at $z=a$, which is given by,
\begin{equation}
E_{c.m.,max}^{2}=\frac{4m^2E^2}{\epsilon}
\label{Ecm2}
\end{equation}

Thus it is seen from the expression above
that the center of mass energy of collision between
ingoing and outgoing particles will be extremely
large if the $\epsilon$, which indicates the deviation
of Kerr metric from extremality, is vanishingly small.
We thus get,
\begin{equation}
\lim_{\epsilon \rightarrow 0} E_{c.m.,max}^{2}=
\frac{4m^2E^2}{\epsilon}\rightarrow \infty
\label{Ecm3}
\end{equation}
This is similar and parallel to what happens in
the blackhole case, where center of mass energy of
collision is divergent only in the limit of approach of
the  Kerr blackhole to extremality,
as described by the BSW mechanism.

\begin{figure}
\begin{center}
\includegraphics[width=0.5\textwidth]{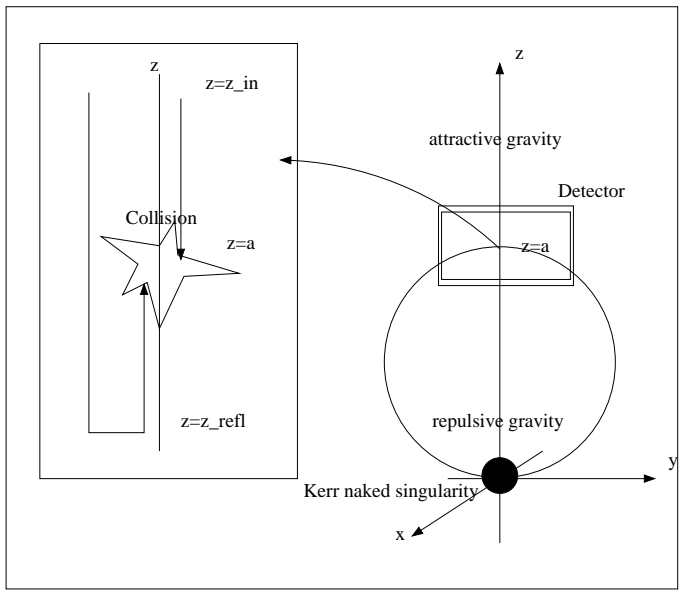}
\caption{\label{fg2}
Schematic diagram of the spatial section of Kerr spacetime
containing a naked singularity. The $z$-axis is the symmetry
axis and the region close to the singularity enclosed in a
circle indicates the repulsive gravity regime. A particle
dropped along $z$-axis from $z=z_{in}$ is reflected
back at $z=z_{refl}=\frac{a^2}{z_{in}}$ and collides with
other ingoing particle at $z=a$. The center of mass energy
of collision is very large. The particle detector is
placed at the point of collision, where the gravity turns
attractive from being repulsive, and it floats
freely in the space.}
\end{center}
\end{figure}

\section{Collider experiment with Kerr naked singularity}

We now describe a collider experiment
which can be used to unravel new physics all the way upto
Planck scale, using the Kerr spacetime with an angular momentum
parameter exceeding mass by a vanishingly small amount,
{\it i.e.} very close to extremality. Fig2 shows a schematic diagram
of the collider experiment in a Kerr geometry.
We consider two identical particles of mass $m$. We can drop the
particles from rest one after the another from a point
$\left(x=0,y=0,z=z_{in}>a\right)$, which
follow  geodesic motion along the $z$-axis which is
the axis of symmetry for the Kerr spacetime. The conserved
energy per unit mass for each particle is $E=\sqrt{f(z_{in})}$.
The first particle initially speeds up when $z>a$
as it falls in.
Its speed $U^{z}$ is maximum when it is at $z=a$,
which is the minimum of the effective potential $f$.
It then slows down and turns back
at $z=\frac{a^2}{z_{in}}$. It speeds up again but now in the
outward direction, speed being maximum at $z=a$
in the outward direction.
We make this particle collide with the second
incoming particle at near $z=a$, when its speed is maximum
in the inwards direction.
The center of mass energy of collision in this process
is given by,
\begin{equation}
E_{c.m.}=\frac{2m\sqrt{f(z_{in})}}{\epsilon}
\label{Ecm2}
\end{equation}
which is arbitrarily large for small enough
values of $\epsilon$. Also, the desired energy of
collision can be obtained or tuned by making an
appropriate choice for the initial point along
the axis $z=z_{in}$ from which the particles are dropped,
thus allowing us to probe new physics at
a range of different very high energy scales.

The particle detector is placed at
$\left(x=0,y=0,z=a\right)$. Since this is a point at the
interface of the attractive and repulsive regimes of
gravity, the detector would stay there at rest on its own,
without the need of any rockets. However some rocket
support might be required
to stabilize its motion along $x$ and $y$ directions.
The measurements from the detector placed
at the site of collision are used to unravel the new physics.
There is no substantial power consumption required
to either accelerate the particles, or maintain
the location of detector
in space, making it a very efficient arrangement
to perform particle collider experiments at arbitrarily
large energies. In contrast, if we want to use the Kerr
blackhole as particle accelerator, much effort and energy
will be needed to stabilize the detector near the
event horizon.

We note, however, that for the collider experiment
suggested above to work in order to probe the Planck scale
physics, the Kerr spin parameter would require an extreme
fine-tuning which has to be very close to unity. If the
colliding particles are neutrons with rest mass
$m_n=940MeV\approx 1GeV$, the center of mass energy of
collision would be comparable to the Planck scale
$E_{pl}\approx 10^{19} GeV$, provided the deviation of spin
from unity is $\epsilon \approx 10^{-38}$. It would be
probably unreasonable to expect an occurrence in nature
of Kerr super-spinning configurations with extreme
fine-tuning of Kerr parameter such as above. Thus it may
not be possible to actually realize such a collider
experiment in practice. However, it still continues to
be an example of a curious Gedanken experiment, the
likes of which over a period of time have played an important
role in the development of relativity theory and
theoretical physics in general.

\section{Comparison with BSW acceleration mechanism}

We now compare these results with BSW particle acceleration
mechanism in the case of extremal or near extremal blackholes
\cite{BSW},\cite{Berti},\cite{Jacobson}.
The BSW mechanism deals with collision between two
infalling particles, which collide near the event horizon
of near extremal Kerr blackhole. Although the horizon is
an infinite blue-shift surface, since infalling particles
arrive almost perpendicularly, their relative velocity
is small. Thus the center of mass energy of collision
would be finite. For divergence of center of mass energy,
the fine-tuning of angular momentum of one of the infalling
particles is necessary. It must have largest possible
angular momentum that still allows it to reach horizon.
This restriction demands that near the horizon,
$\dot{r}=\ddot{r}=0$, where the dot denotes the derivative
with respect to the affine parameter, and $r$ is the
Boyer-Lindquist radial coordinate
\cite{Boyer}.
The condition above implies that the amount of
proper time required for the particle to reach horizon and
participate in collision is infinite. However, in the
case of Kerr naked singularities,
due to the absence of an event horizon and the
transition of nature
of gravity from being attractive to repulsive,
we can consider
the collision between an ingoing and the other
outgoing particles
which have extremely large relative velocity at the
point of collision. Also, since both the conditions
$\dot{z}=0, \ddot{z}=0$ are not realized simultaneously
anywhere along the geodesic, the proper time
required for the collision to happen is finite.


In the BSW mechanism, the maximum center of mass
energy of collision grows as the blackhole approaches
extremality, $-\epsilon=M-a \rightarrow 0$,
because $E_{c.m.,max}^{BSW} \sim \frac{1}{\left(-\epsilon\right)^{1/4}}$.
In our case, because the extremality is approached from
the higher side of the parameter $a$, the maximum center
of mass energy grows twice as fast (as compared to the BSW
mechanism), on a logarithmic scale
$E_{c.m.,max} \sim \frac{1}{\left(\epsilon\right)^{1/2}}$.
Thus to probe Planck scale physics using Kerr blackholes
as particle accelerators, the spin parameter must be tuned
to unity one part in $10^{76}$. Whereas for Kerr naked
singularities the required fine-tuning is brought
down to $10^{38}$.

\section{Possible observational implications of Kerr naked
singularity arising from particle acceleration process}

As we have shown above, the Kerr naked singularities with
spin parameter close to unity provide an environment where
collisions with large center of mass energies can take place.
In a typical astrophysical setting it would be unreasonable
to expect that the fine-tuning necessary for collisions
with center of mass energy comparable to Planck scales can be
actually realized. However, it would be possible to have
collisions with reasonably large center of mass energies
with a reasonable fine-tuning of the spin parameter
close enough to unity as we show below.

We consider a hypothetical situation where a supermassive
astrophysical blackhole candidate, say at the center of the
galaxy, with $M \approx 10^8 M_{\bigodot}$ is modeled by a
Kerr naked singular geometry. Here $M_{\bigodot}=2.10^{30}kg$ is
mass of the sun. Let us assume that the Kerr configuration
under consideration is about one $M_{\bigodot}$ away from
the extremality. That is, upon the addition of one solar mass
the Kerr naked singularity will turn itself into an extremal
Kerr blackhole. In this case we have $\epsilon \approx 10^{-8}$
and the center of mass energy of collision could be as high
as $E_{c.m.}\approx 10^4m$, $m$ being the mass of the colliding
particles. If the colliding particles are assumed to be either
protons or neutrons with mass $m\approx 1GeV$, the center of
mass energy of collisions between them would be
$E_{c.m.}\approx 10^4 GeV \approx 10TeV$. This is the energy
scale at which protons collide at LHC. If the colliding particles
are hypothetical dark matter particles with mass $m\approx 100GeV$,
then the center of mass energy of collision would be
$E_{c.m.}\approx 10^6GeV\approx 10^3TeV$, which is $100$
times larger than the LHC energy scale. The cross-section of
interaction between the particles is dependent on the center
of mass energy of collisions and typically it is large at
the larger values of center of mass energy. It is possible
that new reactions channels for the interactions might also
be available at larger energies. Various high energy particles
with smaller rest masses would be produced in the interactions.
Due to the absence of the event horizon, majority of those
particles might escape to infinity unlike the blackhole
case
where they are absorbed by the event horizon.

Since the galaxies are formed in the dark matter halos, the
central supermassive object, modeled by a Kerr naked singularity
in this case would be surrounded by and accrete the dark matter.
Due to the enhanced density in the high gravity region and
due to the large annihilation cross-section at large center
of mass energy of collisions, dark matter particles would annihilate
to ordinary standard model particles with high energies, majority
of which eventually escape to infinity. These particles could be
possibly detected on earth. It is also quite likely that the high
energy particles would interact with the surrounding gas in the
galaxy and would dissipate their energy as thermal energy. The
heating up of the cloud would then affect the star formation
rate in the surroundings of the galactic center. Thus, if the
Kerr naked singularities exist in nature, they would leave
an imprint on the astrophysical phenomena.

A rigorous analysis will be necessary to come up with
concrete predictions, which will require both the detailed
modeling of the astrophysical scenario as well as the understanding
of the particle physics processes at the required high energies.
This is beyond the scope of the present paper and will be
discussed elsewhere in future.

\section{Concluding remarks}

In this paper we showed that it is possible to have
particle collisions with arbitrarily high center of mass energy
in the vicinity of Kerr naked singularity, with the angular
momentum parameter exceeding unity by a vanishingly small amount,
even if the particles are sent in from infinity from rest.
We note that in this analysis, we have
used the test particle approximation, neglecting the
self-force and backreaction.

We focussed our attention to geodesics
that are restricted along the axis of symmetry. But we also
expect such high energy collisions to take place in the region
around the $z$-axis. The high energy collisions were essentially
a consequence of the fact that the metric coefficient
$g_{zz}^{-1}=f$ is vanishingly small at $z=a$, and because
there is a transition from attractive to repulsive gravity
regime, which allowed us to have collisions between the ingoing
and outgoing particles.

Since, by continuity, both the conditions
hold good in the region nearby the $z$-axis, such collisions
would be realized there as well. We shall present a detailed
analysis elsewhere. Our purpose here has been essentially
to demonstrate the intriguing possibility of having high energy
collisions of particles, for the Kerr spacetime without
an event horizon.

The high energy particles produced in these collisions
could be detected at infinity or these would interact with
the gas in the surroundings. This could in principle lead
to specific observational signatures of the Kerr naked
singularities if they exist in nature.

\end{document}